\documentstyle[prd,aps]{revtex}

\begin{document}
\draft
\title{\bf  Constant mean curvature slices
in the extended Schwarzschild solution and collapse of the lapse. Part II}
\author{ Edward Malec$^{*,**}$ and Niall \'O Murchadha$^{*,+}$}
\address{$^{*}$ ESI, A-1090 Wien, Boltzmangasse 9, Austria.}
\address{ Institute of Physics,  Jagiellonian University,
30-059  Cracow, Reymonta 4, Poland.}
\address{$^{+}$ Physics Department, University College,
Cork, Ireland.}
\maketitle
\begin{abstract}
An explicit CMC Schwarzschildean line element is derived  near the critical point
of the foliation, the lapse is shown to decay exponentially, and the coefficient
 in the exponent is calculated.

\end{abstract}


\subsection{Introduction}

This is a sequel of our previous work on the constant mean curvature (CMC)
slices of the extended Schwarzschild geometry. Here we get CMC foliations
by solving Einstein equations in a particular gauge. A crucial role
is played by a condition (Eq. (\ref{5}) below) imposed on the lapse.
 While this method is completely
equivalent to the other more geometric approach (see \cite{Niall}), it seems
to be more straightforward and technically simpler.  We
focus on the concise derivation of the explicit CMC foliation near the
critical point of the CMC foliation.  The final result is identical to the result
derived in \cite{Niall}.

The  constant mean curvature foliations  have been recently investigated
 numerically in the  simulation of a single spherically-symmetric black hole
 \cite{CMC2}. We hope that our analytic results appear helpful in the 
 verification of the numerical schemes.

\subsection{CMC slicing of the Schwarzschild spacetime}

The notation is the same as in the preceding paper \cite{Niall}. We  define
\begin{equation}
(pR)^2=4\left[ 1 -{2m \over R} + \left( {KR\over 3}- {C\over R^2}
\right)^2\right]  ,
\label{1}
\end{equation}
\begin{equation}
\gamma(R,t) =1+ 8\partial_tC\int_{R}^{\infty }dr{1\over  r^5p^3}.
\label{2}
\end{equation}
and
\begin{equation}
N =\gamma {pR\over 2}.
\label{3}
\end{equation}
Here $m$ is    the  mass, $K$ (the trace of the extrinsic curvature)
is a constant    and C is a   time-dependent parameter which measures the transverse part
of the extrinsic curvature.

The Schwarzschild line element, expressed in terms of coordinates adapted to
the constant mean curvature foliation, is given by   \cite{Iriondo}
\begin{eqnarray}
ds^2&=&-dt^2\Biggl( N^2 -\gamma^2\Bigl( {KR\over 3}- {C\over
R^2}\Bigr)^2\Biggr)
+4N {{C\over R^3}-{K\over 3}\over p^2R}dtdR+ {4\over (pR)^2}dR^2+
R^2d\Omega^2.
\label{4}
\end{eqnarray}
The hypersurfaces of constant time are CMC slices,  asymptotic to the CMC slices of
Minkowskian geometry.

\subsection{Elliptic slicing condition }

A minimal surface is a locus of points  defined by the condition
$p =0$. Choose a CMC Cauchy hypersurface $\Sigma_C$
of the extended Schwarzschild manifold   corresponding to a parameter $C$
 and let $R_0$ be an areal radius corresponding to a
 simple zero  of $p^2$; that is $p^2(R_0)=0$ but
 $\partial_{R}p^2|_{R_0}\ne 0$. Futhermore, assume that
\begin{equation}
{\partial_rN \over \sqrt{a}}|_{R_0}=0
\label{5}
\end{equation}
at   $R_0$. 
The condition (\ref{5}) yields  
\begin{equation}
 \partial_tC={1\over 8I(R_0)}.
\label{6}
\end{equation}
Here
\begin{equation}
I(R_0)\equiv  \int_{R_0}{dr\over pr}{6{C^2\over r^4}+{K^2r^2\over 3}\over
\Biggl( 2m+{2KC\over 3}+{2K^2r^2\over 9}  -{4C^2\over r^3} \Biggr)^2}.
\label{7}
\end{equation}
The value of the lapse function $N$  at the minimal surface,
that is at the areal  radius $R_0$, can be shown to be equal (using Eqs.
(\ref{1} -- \ref{3})) to
\begin{equation}
N = {dC\over dt} {1\over m+{KC\over 3} +{K^2R^3_0\over 9}-2{C^2\over R^3_0}}.
\label{8}
\end{equation}
The lapse $N $ is strictly positive at the minimal surface corresponding to a simple
zero $R_0$. Eqs. (\ref{1} -- \ref{3}) imply that $N (R)> 
N (R_0)$ if $R>R_0$
 and therefore the lapse exists on all of $\Sigma_C$.
Equation (\ref{6}) dictates the rate of change of the parameter $C$.
It is clear that one
can uniquely construct  a foliation of a part of  the extended  Schwarzschild
geometry by imposing the condition (\ref{5}) at minimal surfaces 
on all slices to the future of a given one. The leaves of the resulting foliation  
connect two null infinities of the extended Schwarzschild spacetime.
This gives us  a curve $R_0(t)$ of zeroes of the mean curvature $p$.
It is evident, just by inspecting the explicit solution presented above,
that the line running along the locations of minimal surfaces $R_0(t)$   
can be arranged to be smooth. It can be chosen to coincide with the `vertical'
 $t = 0$ axis in standard Schwarzschild coordinates.

This construction breaks down when $R_0$ ceases to be a simple zero of $p^2$,
since expressions appearing in Eqs. (\ref{6}) and (\ref{8}) become unboundedly large.
The   goal of this paper is to show the asymptotic behaviour
of the lapse at the critical minimal surface.

\subsection{The  evolution of C near critical point}

Let $C_*$ and $R_*$ be degenerate, that is  such that the  zero of $p^2$
ceases to be
simple. In this  case both $p$ and its derivative $\partial_Rp$ vanish;
that means
that
\begin{eqnarray}
&& 1- {2m\over R_*} -{2KC_*\over 3R_*}+{K^2R^2_*\over 9}+{C^2\over R^4_*}=0,
\nonumber\\
&&
2m +{2KC_*\over 3} + {2K^2R^3_*\over 9} -{4C^2_*\over R^3_*}=0.
\label{9}
\end{eqnarray}
One can easily show, if $C_*$ and $R_*$ are critical, then
the sign of
\begin{equation}
\beta  \equiv -2C_*+{2\over 3}KR^3_*
\label{10}
\end{equation}
is the same as the sign of $-C_*$.

There  exists critical values  of  $C_*$  that are positive ($C_*^+$)
or negative  ($C_{*-}$).
For definiteness we shall consider only the case when $C(t=0) > C_{*-}$,
therefore the only  limiting case we consider is that with $C\rightarrow  C_*^+$.
(That choice  corresponds to    a foliation formed by leaves
connecting two null infinities which moves forward in time - see a discussion in 
Sec IV in  \cite{Niall}).
For simplicity we will drop the $^+$ suffix and $C_*$ will mean
a positive   critical parameter.  From the dynamical equation
(\ref6) follows that $C$ can only increase.

Next, let us introduce the notation that
\begin{eqnarray}
&&\epsilon\equiv C_*-C  \nonumber\\
&& R_0\equiv R_* +\delta .
\label{11}
\end{eqnarray}
where both $\delta $ and  $\epsilon $  are positive and small.

The equation $p(R_0)=0$ yields a nonlinear algebraic equation whose truncation
gives
\begin{equation}
\delta^2 A  +\epsilon \beta   =0.
\label{12}
\end{equation}
Here $A \equiv 2R^2_*+K^2R^4_*$.
Eq. (\ref{12}) is in fact the    Lyapunov - Schmidt   reduced equation
constructed
according to the standard rules \cite{Trenogin}.
Therefore in the vicinity of the critical point we have
\begin{equation}
\delta =\sqrt{-\beta \epsilon \over A}.
\label{13}
\end{equation}
The function $p$ can be expressed in a form
\begin{equation}
{pr\over 2} =\sqrt{1-{R_0\over r}}\Biggl[ {\kappa \delta \over R_0}+
{K^2\over 9}(rR_0+r^2-2R_0^2 )-{C^2\over R^4_0} ({R_0\over r} +   {R_0^2\over r^2}
+{R_0^3\over r^3}-3)\Biggr]^{1/2}.
\label{p}
\end{equation}
The insertion of (\ref{11}), (\ref{13}) and (\ref{p}) into the equation
(\ref{7}) and the change of the integration variable to $y=\sqrt{1-{R_0\over
r}}$, yield after a simple but tedious algebra
\begin{equation}
I(R_0)\approx \sqrt{R_*}\int_0^1dy{F_1\over
\sqrt{ \kappa \delta +y^2F_2}( \kappa \delta
+y^2F_3)^2}.
\label{14}
\end{equation}
Here  the functions $F_1, F_2$ and $F_3$ are given by
\begin{eqnarray}
F_1(y)&=&{K^2R_*^3\over 3(1-y^2)^4}+{6C^2_*\over R^3_*}(1-y^2)^2,
\nonumber \\
F_2(y)&=&  {K^2R^3_*(3-2y^2)\over 9(1-y^2)^2}+{C^2_*\over R_*^3}
\Bigl( 6-4y^2+y^4\Bigr) ,
\nonumber \\
F_3(y)&=& (3-3y^2+y^4)\Bigl( {2K^2R^3_*\over 9(1-y^2)^3}
+{4C^2_*\over R^3_*}\Bigr) ,
\label{15}
\end{eqnarray}
while $\kappa $ reads
\begin{equation}
\kappa = {2K^2R^2_*\over 3}+{12C^2_*\over R^4_*}.
\label{16}
\end{equation}

\subsection{Limiting behaviour of the foliation.}

The asymtotic behaviour of $C$ will be dominated by the $1/\delta^2$ part
of $I(R_0)$. As will be shown later, $C$ tends exponentially to $C_*$; the
attenuation factor in the exponent depends only on the leading term of $I(R_0)$.
It is useful to define $z=y /\sqrt{\kappa \delta }$. Then one obtains
$I(R_0)={\sqrt{R_*}\over \kappa^2\delta^2}\times I_d$, where
\begin{equation}
I_d\equiv \int_0^{1/\sqrt{\kappa \delta }}dz{F_1(\sqrt{\kappa \delta}z)\over
\sqrt{  1 +{z^2\over R_*}F_2(\sqrt{\kappa \delta }z) }( 1
+{z^2\over R_*}F_3(\sqrt{\kappa \delta }z))^2}.
\label{17}
\end{equation}
One can split the integral $\int_0^{1/\sqrt{\kappa \delta }}$ into
two parts:  $\int_0^{1/\sqrt{\kappa \delta }}=
\int_0^{1/\sqrt{10^4\kappa \delta }}+
\int_{1/\sqrt{10^4\kappa \delta }}^{1/\sqrt{\kappa \delta }}$.
It is easy to check that the contribution coming from the second
integral goes to zero as $\delta $ approaches zero. Therefore $F_1\approx R_*\kappa /2$,
$F_2\approx R_*\kappa /2$ and $F_3\approx R_*\kappa $. Thus the  first
integral (and also the integral $I_d$)  is well approximated by
\begin{eqnarray}
I&=&{\kappa R_*\over 2}
 \int_0^{\infty } dz {1\over \sqrt{1+ {R_*\kappa \over 2}z^2}(1+R_*\kappa z^2)^2}=   \nonumber \\
&& {\sqrt{\kappa R_*}\over 2}
 \int_0^{\infty } dz {1\over \sqrt{1+ {z^2\over 2}}(1+z^2)^2}.
\label{18}
\end{eqnarray}
The integral $I_z=\int_0^{\infty } dz {1\over \sqrt{1+ {z^2\over 2}}(1+z^2 )^2}$
 can be explicitly evaluated and gives $\sqrt{2}/2$.

In summary, near the critical point we have
\begin{eqnarray}
I(R_0) &=&{\sqrt{R_*}\over \kappa^2\delta^2}I \approx
\nonumber \\
&&
 {\sqrt{2}\over 4\epsilon } {R_*A \over \kappa^{3/2}|\beta |}
\label{19}
\end{eqnarray}
The insertion of    (\ref{11})   and  (\ref{19})
into (\ref{6}) yields
\begin{equation}
\partial_t\epsilon  = - \Gamma \epsilon  ,
\label{20}
\end{equation}
where
\begin{equation}
\Gamma = {|\beta |\kappa^{3/2}  \over 2\sqrt{2} A  R_*}.
\label{21}
\end{equation}
Eq. (\ref{20}) immediately implies that $\epsilon $
 approaches 0    exponentially as
\begin{equation}
\epsilon (t) =\epsilon_0e^{-t\Gamma },
\label{23}
\end{equation}
where $\epsilon_0 $ is an initial value of the parameter. Taking into
account  relations (\ref{11}) and (\ref{13}), one  can
conclude that the parameter $C$ and the minimal radius $R_0$
tend exponentially to their critical values $C_*$ and $R_*$,
respectively.

Finally, collecting the above information and putting it
into Eq. (\ref{8}), we obtain the asymptotic behaviour of the
lapse function near the critical point
\begin{equation}
N = N_0 e^{-t\Gamma /2}.
\label{24}
\end{equation}
This is exactly the same result as that obtained in \cite{Niall};
in order to show equivalence, use expressions for the extrinsic curvatures
(valid in the upper quadrant of the extended Schwarzschild geometry),
 which imply $K=(2mR_*-3m)/\sqrt{2mR^3_*-R^4_*}$
and $C_*=(3mR^3_*-R^4_*)/  \sqrt{2mR^3_*-R^4_*}$.
In the case of maximal slicing ($K =0$) the decay constant $\Gamma /2$
is equal to $4/(3\sqrt{6})$, in   agreement  with the  analytic derivation of \cite{MAX}
and close to the numerical result of \cite{CMC1a} .
The asymptotic behaviour of
$\gamma $ and $p$ in a region close to the line $R_0$ is given by
\begin{eqnarray}
\gamma &=& \gamma_0
 {e^{-{t\Gamma \over 4}}\over \sqrt{1-{R_0\over r}}}\nonumber \\
{pr\over  2}&=&p_0
 \sqrt{1-{R_0\over r}}  e^{-t\Gamma \over 4}
 \label{25}
\end{eqnarray}
Here $\gamma_0$ and $p_0$ are initial values of $\gamma   \sqrt{1-{R_*\over r}}$
and $pr/(2\sqrt{1-{R_*\over r}})$, respectively. The four constants ($\epsilon_0, N_0,
\gamma_0$ and $p_0$) can be expressed in terms of one free parameter (say, $\epsilon_0$)
and $A, \beta, \kappa $.
Equations (\ref{3}), (\ref{14}), (\ref{23}) - (\ref{25}) and    $C=C_*-\epsilon_0e^{-t\Gamma }$
suffice to construct the metric (\ref{4}) near the line $R_0(t)$ of
minimal surfaces.

Acknowledgments.    This work has been suported
in part  by the KBN grant 2 PO3B  006 23.

\end{document}